\documentclass[aps,prl,amsmath,amssymb,reprint]{revtex4-1}

\usepackage{graphicx}
\usepackage{dcolumn}
\usepackage{bm}
\usepackage{color}
\usepackage[table]{xcolor}

\begin{document}

\draft
\title{Persistence of slow fluctuations  in the overdoped regime of Ba(Fe$_{1-x}$Rh$_{x}$)$_2$As$_2$ superconductors}

\author{L. Bossoni,$^{1,2}$ M. Moroni$^{1}$, M. H. Julien,$^{3}$ H. Mayaffre,$^{3}$ P. C. Canfield,$^{4}$A. Reyes,$^{5}$ W. P. Halperin,$^{6}$ and P. Carretta$^{1}$}

\address{$^{1}$ Department of Physics, University of Pavia-CNISM, I-27100 Pavia, Italy}
\address{$^{2}$ Huygens-Kamerlingh Onnes Laboratory, Leiden University, 2333CA Leiden, The Netherlands}
\address{$^{3}$ Laboratoire National des Champs Magnetiques Intenses, CNRS - Universit\'e Grenoble Alpes - EMFL, 38042 Grenoble, France}
\address{$^{4}$ Ames Laboratory US DOE and Department of Physics and
Astronomy, Iowa State University, Ames, Iowa 50011, USA}
\address{$^{5}$ National High Magnetic Field Laboratory, Tallahassee, FL 32310, USA}
\address{$^{6}$ Department of Physics and Astronomy, Northwestern University, Evanston, IL 60208, USA}

\begin{abstract}
We present nuclear magnetic resonance evidence that very slow
($\leq 1$ MHz) spin fluctuations persist into the overdoped regime
of Ba(Fe$_{1-x}$Rh$_{x}$)$_2$As$_2$ superconductors. Measurements
of the $^{75}$As spin echo decay rate, obtained both with Hahn
Echo and Carr Purcell Meiboom Gill pulse sequences, show that the
slowing down of spin fluctuations can be described by short-range
diffusive dynamics, likely involving domain walls motions
separating $(\pi/a,0)$ from $(0,\pi/a)$ correlated regions. This
slowing down of the fluctuations is weakly sensitive to the
external magnetic field and, although fading away with doping, it
extends deeply into the overdoped regime. 
\end{abstract}

\maketitle

\draft

\narrowtext
\section{Introduction}

The understanding of the electronic properties of iron-based
superconductors has significantly progressed over the last years.
Superconductivity arises on the verge of an ordered magnetic phase
with wavevector $(\pi/a,0)$ (or $(0,\pi/a)$), characterized by an
orthorhombic distortion and by a population imbalance between $d_{xz}$
and $d_{yz}$ Fe orbitals \cite{Chubukov2015}. Most of the debate
is now focused on determining how the lattice, spin and orbital
degrees-of-freedom intertwine \citep{FernandesNP}. Nonetheless, a
detailed comprehension of the spin dynamics, that are widely
thought to play a central role, is still lacking. In particular,
nuclear magnetic resonance (NMR) experiments have suggested that
the sharp magnetic transition at low electron doping evolves into
a cluster spin-glass behavior near the optimal doping level for
superconductivity \cite{Dioguardi2013}. More recent NMR and
neutron scattering studies \cite{Hu2015,Lu2014} gave further
support for the presence of a cluster spin-glass phase coexisting
with superconductivity at low electron doping. Remarkably, even
when long-range magnetic order and cluster spin-glass phases
vanish, enhanced low-frequency fluctuations (MHz range) persist in
the normal phase of different families of iron-based
superconductors \cite{Bossoni2013,
Dioguardi2013,Dioguardi2015,Hammerath2015,Hammerath2013,Oh2011}.
The origin of these slow dynamics still remains unsettled.

In general, spin dynamics may become glassy (i.e. slow and
inhomogeneous) under the influence of quenched disorder or in the
case of competing interactions \cite{Schmalian2000}. In
iron-pnictides, slow fluctuations have been argued to arise from
the motion of domain walls \cite{Mazin2009,Xiao2012,CurroNJP} that
separate $(\pi/a,0)$ and $(0,\pi/a)$ correlated spin fluctuations, a
situation analogous to that observed in frustrated vanadates
\cite{Melzi}. The slowing down of domain wall fluctuations may be
related to pinning driven by quenched disorder \cite{Dioguardi2015} or
might be intrinsically due to underlying geometric frustration and
long-range Coulomb repulsion \cite{Mahmoudian2015}. A recent
theory also argues that phase separation could drive a glass-like
freezing \cite{Wang2015}, but this implicitly requires the onset
of superconductivity whereas low frequency fluctuations are
observed to develop already in the normal state.
\newline
While the amplitude of slow dynamics must increase on approaching
the spin density wave (SDW) transition occurring at low electron doping, it
is not yet clear what happens on moving towards the overdoped
superconducting regime. In this article, we show from NMR echo
decay measurements that very slow spin fluctuations actually
persist at least up to 11\% doping in the overdoped regime of
Ba(Fe$_{1-x}$Rh$_x$)$_2$As$_2$. Rh doping induces an increase of
electron concentration in the conduction bands of BaFe$_2$As$_2$
(Ba122) very much akin to Co doping. In fact, Rh and Co-doped
Ba122 display practically identical phase diagrams \cite{Canfield,Kim2016}.

By combining different spin-echo techniques, we demonstrate that
the low-temperature increase in the transverse relaxation rate
$1/T_2$ originates from an activated slowing down of the
fluctuations, rather than from an increase of their amplitude.
Moreover, it is shown that the activated correlation time
describing $1/T_2$, with an energy barrier decreasing with Rh
doping, accounts also for the behavior of the spin-lattice
relaxation rate $1/T_1$. We finally evidence that this type of
fluctuations extends at least up to about 11\% of doping,
differently from earlier results based on $1/T_1$ analysis
\citep{Dioguardi2013}, suggesting that the vanishing of the
superconducting phase in Ba(Fe$_{1-x}$Rh$_x$)$_2$As$_2$ is
followed by the concomitant fading of these low-frequency
excitations.

\section{Materials and Methods}

$^{75}$As NMR experiments were performed on Ba(Fe$_{1-x}$Rh$_x$)$_2$As$_2$
single crystals \cite{Ni2008} with Rh content and superconducting
transition temperature ($T_c$) of $x=4.1$\% ($T_c=13.6$ K), $x
=6.8$\% ($T_c=22.4$ K), $x= 9.4$\% ($T_c=15.1$ K) and $x = 10.7$\%
($T_c=12.25$ K), respectively. $T_c$ was determined by
superconducting quantum interference device (SQUID) magnetometry
prior to the NMR experiment and also checked \textit{in situ}, via
the observation of the detuning of the NMR tank circuit. The
magnetic field $\mathbf{H}$ was applied along the crystallographic
$c$ axis, unless otherwise stated.

\begin{figure}[htbp]
    \centering
    \includegraphics[width=7.5cm,keepaspectratio]{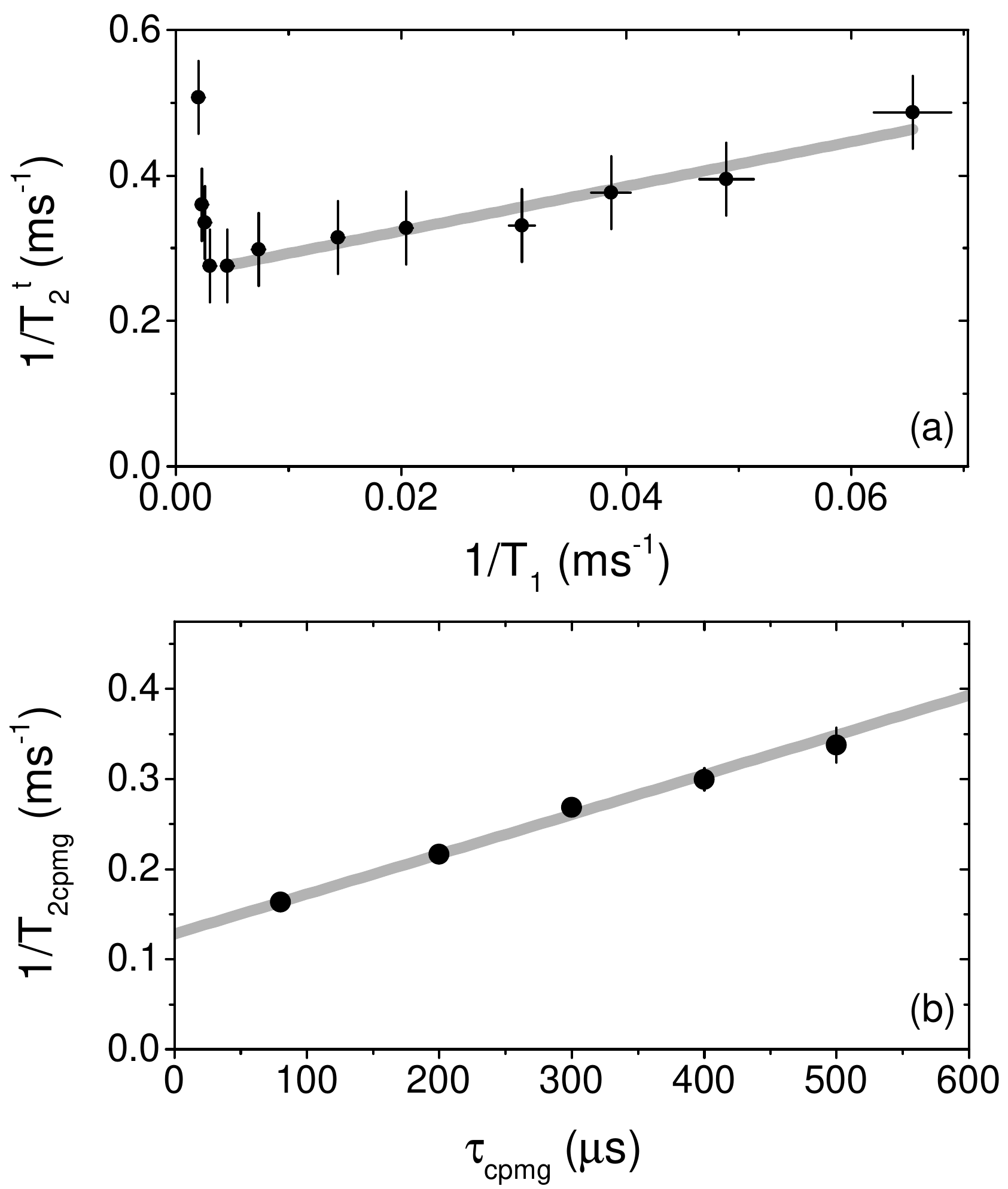}
    \caption{\textbf{(a) }The raw Hahn echo decay rate ($T_2^t$) versus the spin-lattice relaxation rate in the normal phase of the
    $x=9.4$ \% sample at $H= 6.4$ T. $T_2^t$ is defined as the time at which the normalized echo amplitude decays to $1/e$.
    The temperature is an implicit parameter. The grey line is the linear fit to the data above 20 K.
 \textbf{(b)} $1/T_{2cpmg}$ plotted as a function of $\tau_{cpmg}$, for the compound $x=6.8\%$,
 measured at 6.4 T, and 70 K. The grey line is a fit to extrapolate the intrinsic $T_{2cpmg}$ value. }
    \label{nFig1}
    \end{figure}

\begin{figure}[htbp]
    \centering
    \includegraphics[width=8.5cm,keepaspectratio]{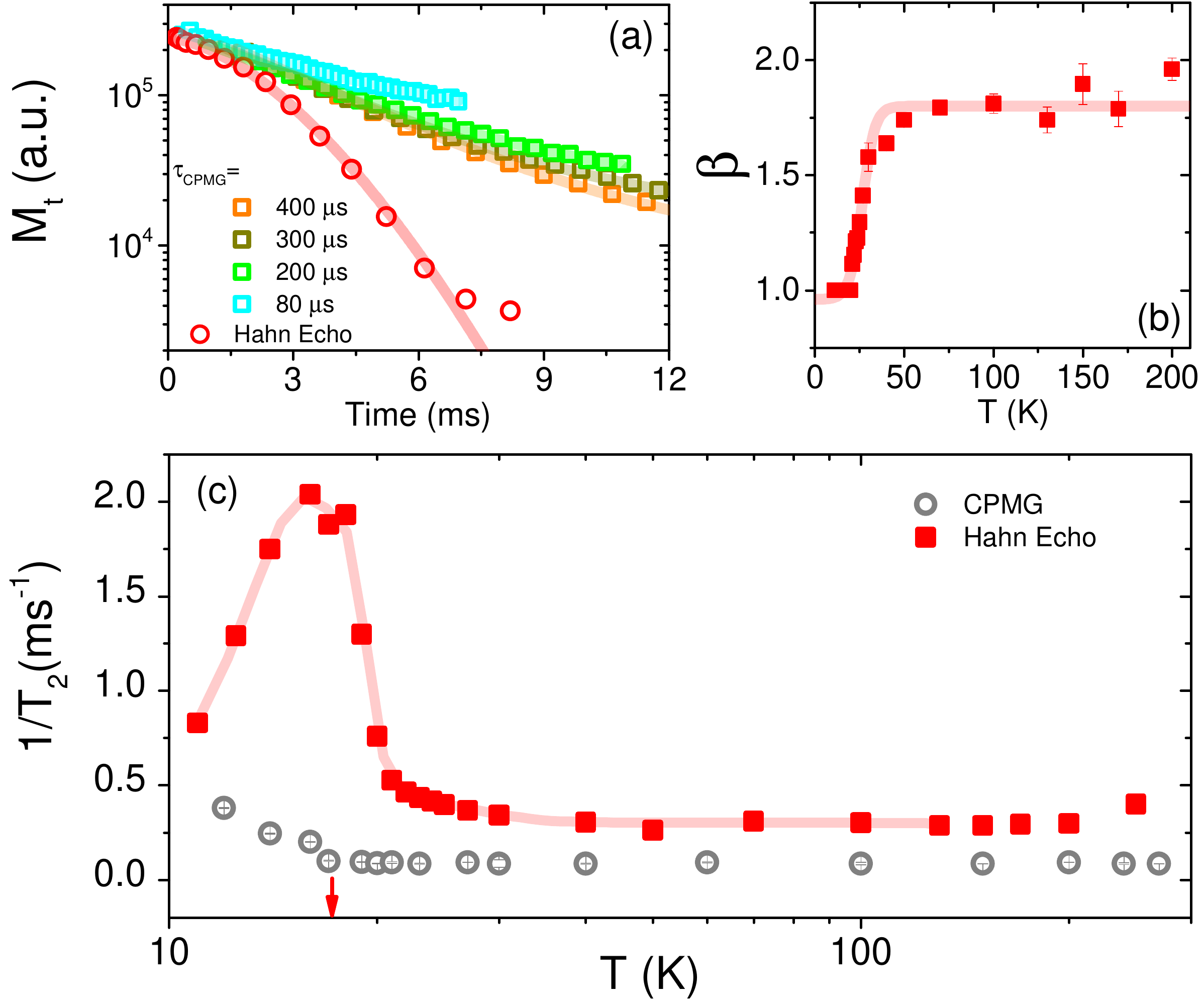}
    \caption{\textbf{(a)} Hahn (red circles) and CPMG echo decays, measured on the x=6.8\% compound, at $T= 70$~ K and
    $H= 6.4$ T. The delay $\tau_{cpmg}$ between CPMG echoes is indicated. The comparison between the two pulse sequences evidences the deviation from single exponential in the Hahn echo decay. The solid lines are the fit to the equations discussed above.
    \textbf{(b)} The stretched exponent is reported as a function of the temperature, measured for the same sample, at 11 T. The shadowed line is a guide to the eye.
\textbf{(c)} Hahn (red squares) and CPMG (grey circles) echo decay
rates, as a function of the temperature, measured in a magnetic
field of 11 T. The raw data have been corrected for the Redfield
term. The arrow marks T$_c$. The red line is a guide for the eye.
 }
    \label{nFig2}
    \end{figure}

The echo-decay time was first measured by the standard Hahn echo
sequence: $\pi/2$ - $\tau$ - $\pi$ \cite{Hahn}. Since the
spin-lattice relaxation time $T_1$ and the raw Hahn echo decay
time have both values in the 1-100 ms range, one can expect a
sizable contribution of $T_1$ processes to the echo
decay (Redfield term \cite{Redfield1968,Slichter}), as confirmed by
the linear dependence of the raw echo decay rate $1/T_2^t$ on
$1/T_1$ at high temperature (Fig. \ref{nFig1} \textbf{(a)}).
In fact, the Hahn echo decay $M_t(2\tau)$ can be written
\cite{WC1995}:
\begin{equation}
M_t(2\tau)=M(2 \tau)\exp\left( -\frac{2\tau}{T_{1R}}\right)
\end{equation}
evidencing that the relaxation involves both spin-lattice
relaxation processes, via the $T_{1R}$ term, and a $T_1$
independent $M(2\tau)$ term. In case of an anisotropic
spin-lattice relaxation rate, Walstedt and
coworkers \cite{WC1995} obtained a general result for the central
($\frac{1}{2}\rightarrow -\frac{1}{2}$) transition of half
integer spin, which for I=3/2 is:

\begin{equation}
\frac{1}{T_{1R}^{\parallel}}=\frac{3}{T_{1}^{\parallel}}+\frac{1}{T_{1}^{\perp
}}
\end{equation}
where the symbols $\parallel$ and $\perp$ refer to the magnetic
field orientation with respect to the crystallographic $c$ axis.
Once the raw echo decay data have been corrected for the
spin-lattice relaxation term, the Hahn echo decay contribution
$M(2\tau)$ was analyzed. The Hahn echo decay was found to deviate
from a single exponential (Fig. \ref{nFig2}\textbf{(a)}) and could
be fit in general to a stretched exponential, $M(2\tau)=M_0
\exp(-(2\tau/T_{2})^{\beta})$, where the stretched exponent showed a marked temperature dependence ( Fig. \ref{nFig2}\textbf{(b)}).

The echo decay time was also measured with the
Carr-Purcell-Meiboom-Gill (CPMG) sequence \cite{CPMG}, in which
the $\pi/2$ pulse is followed by a comb of $\pi$ pulses,
separated by a time $\tau_{cpmg}$, ranging between 80 $\mu s$ and
500 $\mu s$ (Fig. \ref{nFig2}\textbf{(a)}). The echo amplitude
decays exponentially with time and the decay rate is found to increase
linearly with $\tau_{cpmg}$ (Fig. \ref{nFig1}\textbf{(b)}).
Hence, $1/T_{2cpmg}$ can be conveniently  defined by taking the
value extrapolated for $\tau_{cmpg}\rightarrow 0$.

In the iron-based superconductors, few works have attempted to measure the spin-echo decay time also
with a CPMG sequence 
\cite{Mukhopadhyay2009,Oh2011}. However, a comparative study
between Hahn and CPMG sequences is beneficial in revealing the
presence of slow spin dynamics. At high temperature ($T \geq
50$~K), the spin-echo decay measured by both methods is
temperature independent, with $1/T_{2cpmg}< 1/T_2$ (Fig.
\ref{nFig2}\textbf{(c)}), and smaller than the value $1/T_2^{dip}=
1.4 $~ms$^{-1}$ expected from the dipolar interaction between As
nuclei \cite{Bossoni2013}. The origin of this discrepancy will be
discussed subsequently. Since the $\pi$ pulses of the CPMG
sequence were not phase alternated, $T_{2cpmg}$ could be affected
by spin-locking effects \cite{Borsa1,Borsa2,SL}. This could
explain the difference between $T_{2}$ and $T_{2cpmg}$ at high
temperature where both times are $T$ independent, but it does not affect
our conclusions concerning the different T-dependence observed at
low temperature.

\begin{figure}[htbp]
    \centering
    \includegraphics[width=9cm,keepaspectratio]{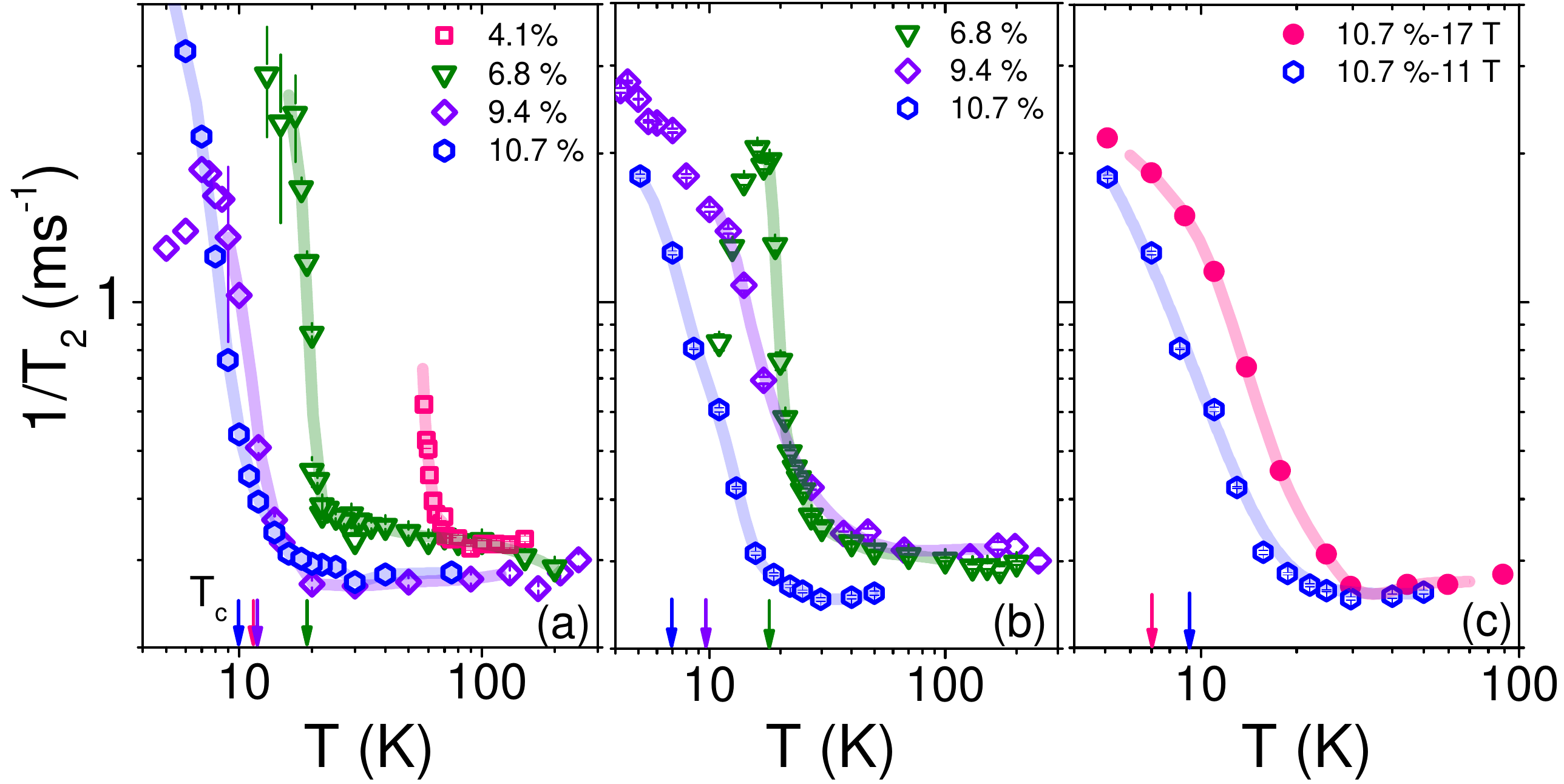}
    \caption{Hahn Echo decay rates as a function of the temperature, for Ba(Fe$_{1-x}$Rh$_x$)$_2$As$_2$ for different magnetic fields:
    \textbf{(a)} H=6.4 T,  \textbf{(b)} H=11 T and \textbf{(c)} for x=10.7\% at 11 and 17 T. The arrows mark T$_c$.
    The shadowed lines are guides to the eye. }
    \label{nFig3}
    \end{figure}

Remarkably, $1/T_2$, measured by the Hahn
echo sequence, shows a pronounced enhancement starting above $T_c$, on cooling (Fig. \ref{nFig2} \textbf{(c)}). This increase, observed 
at all magnetic fields (Fig. \ref{nFig3}), is not detected
by the CPMG echo sequence (Fig. \ref{nFig2}\textbf{(c)}). This
dichotomy is observed in all the studied samples, thus
corroborating and complementing the initial findings of
Ref.~\cite{Bossoni2013}.

\begin{figure}[htpb!]
\centering
\includegraphics[height=8cm, keepaspectratio]{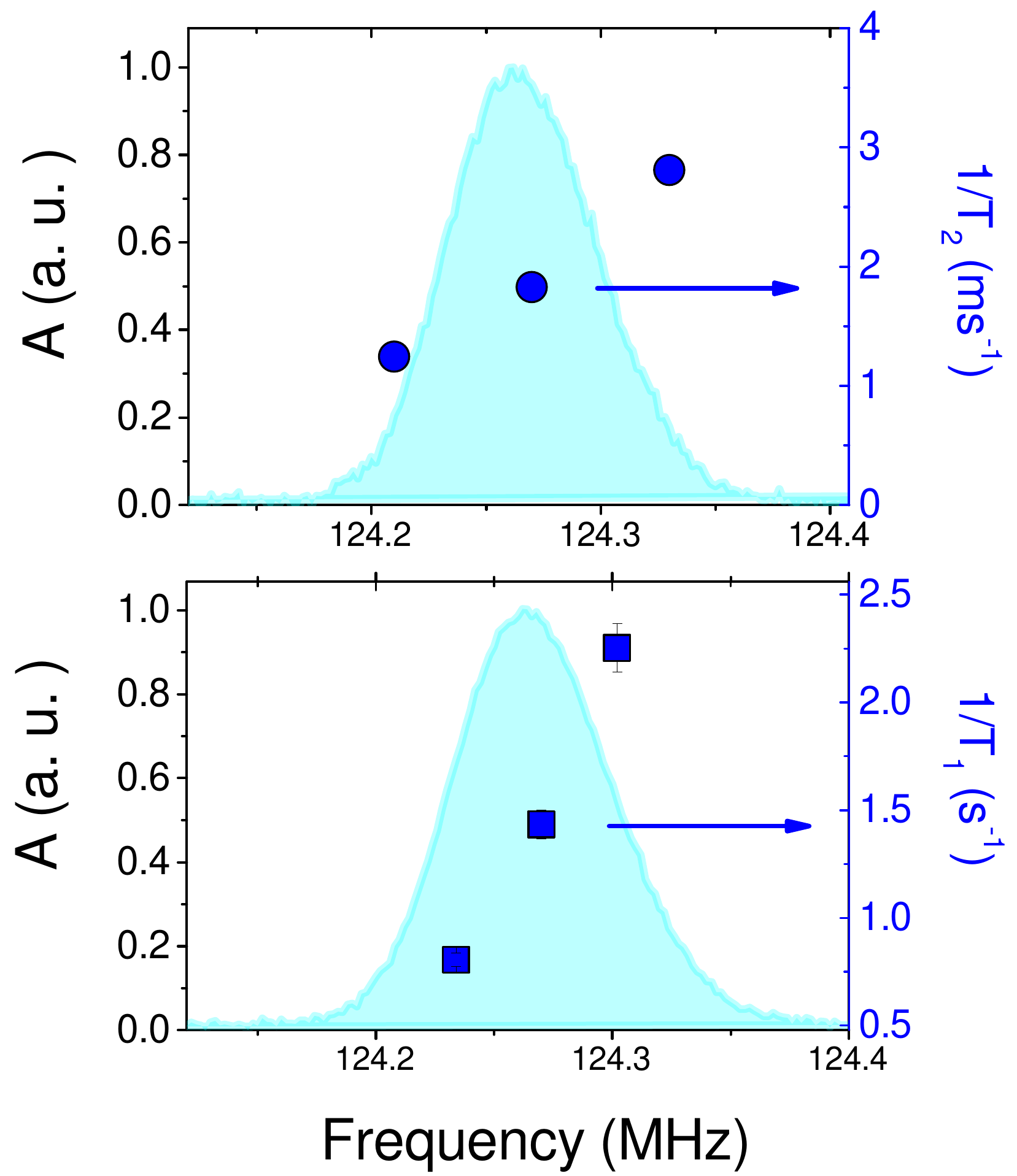}
\caption{\textbf{(top)} $1/T_{2}$ (right axis) as a function of the frequency, measured at 17 T and 150 K, for the $x=10.7\%$ sample. \textbf{(bottom)} $1/T_{1}$ (right axis) as a function of the frequency, measured at 17 T and 7 K, for the same sample. The left axes represent the spectral intensity, in arbitrary units.}
\label{nFig4}
\end{figure}

Finally, we point out that the spin-lattice and transverse
relaxation times are not homogeneous across the NMR line. The
results in Fig. \ref{nFig4} show two representative plots for the
overdoped sample with $x=10.7\%$, at 17 T. The $T_2$ and $T_1$
variation is $\sim$ 65 \%, across the whole spectrum. This
spectral distribution of relaxation times suggests that not all
spins have the same spin temperature. The values of $1/T_1$ and
$1/T_2$ reported in this manuscript were recorded irradiating the
central part of the spectrum. We notice that the same spatial
magnetic inhomogeneity was also observed in the T$_1$ measurements
of the Co-doped compounds \cite{Ning2009,Dioguardi2015}.

\section{Discussion}

Let us first consider the difference between the Hahn and CPMG
sequences. While the former is very effective in refocusing the
dephasing of the in-plane nuclear magnetization coming from
\textit{static} field inhomogeneities, the amplitude of the Hahn
echo is decreased by dynamics with a fluctuation time scale of the
order of the separation between the $\pi/2$ and $\pi$ pulses. In
case of diffusive-like dynamics in a field gradient
\textbf{$\nabla\mathbf{B}$}, described by a diffusion coefficient
$D$, one should weight $M(2\tau)$ by $ \sim
\exp(-\gamma^2|\nabla \mathbf{B}|^2D(2\tau)^3)$. Carr and Purcell (CP) \cite{CPur} devised a way to
quench the effects of these dynamics by slicing the time $\tau$
over which the dynamics would irreversibly quench the echo
amplitude with many $\pi$ pulses, separated by $\tau_{cpmg}\ll
\tau$.
Accordingly, the dynamics would become effective in reducing the echo amplitude only if its
characteristic time scale is of the order of $\tau_{cpmg}$. The
original CP sequence was later implemented into the CPMG one, in order to avoid
phase error accumulation.

Now, in Rh-doped Ba122 compounds we observed a linear increase of
$1/T_{2cpmg}$ with $\tau_{cpmg}$, which is typical of systems,
such as platinum nanoparticles \cite{Yu1993}, where restricted
electron spin diffusion in a non-uniform magnetic field occurs
\cite{Robertson1966, Slichter,Mukhopadhyay2009,Oh2011}. In the
$\tau_{cpmg}\rightarrow 0$ limit, $1/T_{2cpmg}$ is no longer
affected by the dynamics and only the irreversible decay due to
nuclear dipole-dipole interaction between $^{75}$As nuclei should
be effective. This intrinsic decay time, $T_{2i} \sim $ 10 ms,
should be compared with the much shorter one estimated from
lattice sums for $^{75}$As-$^{75}$As dipolar interaction, equal to
0.7 ms. The long experimental value of $T_{2i}$ suggests that not
all the As nuclei are contributing to the dipolar field
distribution, as in presence of a mechanism quenching the nuclear
spin flip-flop mechanism \cite{Slichter}. The suppression of the
latter occurs when the inhomogeneous NMR linewidth is much larger than the
dipolar coupling between $^{75}$As nuclei, as justified below. The
quenching of flip-flop mechanisms is further supported by the
distribution of relaxation rates observed across the NMR line
(Fig. \ref{nFig4}), indicating the absence of a common
spin-temperature among $^{75}$As nuclei and suggesting that the
electronic system is highly inhomogeneous.

Unlike $1/T_{2cpmg}$ $(\tau_{cpmg}\rightarrow 0)$, the Hahn echo decay
rate $1/T_2$ is sensitive to electron spin diffusive dynamics.
Given the inhomogeneous nature of the electronic texture
\cite{Julien2009,Laplace2012}, an internal magnetic field gradient
\textbf{$\nabla\mathbf{B}$} could originate from a spatial
inhomogeneity of the spin susceptibility $\Delta\chi$ or,
equivalently, of the local magnetization $\Delta \chi H$
\cite{Mitchell2010,Brown1993}. Hence, $\nabla\mathbf{B}\simeq
\Delta\chi \mathbf{H}/2a$, where $2a$ defines a typical domain
size \cite{Mitchell2010}. A successful approach to treat the echo
relaxation in the case of restricted diffusion was presented by
Robertson \cite{Robertson1966,Wayne1966}. Robertson showed that it
is possible to describe restricted spin diffusion by an equivalent
mechanism of unrestricted diffusion in a periodic field gradient.
As mentioned above, we assume that the source of \textit{internal} field
inhomogeneity here comes from the distribution of hyperfine fields
at $^{75}$As nuclei, affecting the NMR linewidth $\Delta\nu$.
Thus, we can write the field gradient probed by the nuclei as
\begin{equation}
\nabla\mathbf{B}_{hyp} = \frac{\pi\Delta\nu }{ a \gamma}
\end{equation}
where $\gamma$ is the nuclear gyromagnetic ratio. From the equation above, by taking a linewidth of 30 kHz \cite{Bossoni2013}, and $2a$ equal to few lattice steps \cite{Millis2014,Milan2013}, the internal field gradient results $\nabla B \sim 10^8$ G$/$cm.
Therefore two As nuclei separated by 0.6 nm experience
a Larmor frequency difference of about 4 kHz, which is much
larger than the dipolar interaction, estimated from lattice sums to be $\sim$ 200 Hz. This justifies the fact that nuclear spin flip-flop processes are quenched \cite{GR}.

\begin{figure}[htbp]
    \centering
    \includegraphics[width=8cm,keepaspectratio]{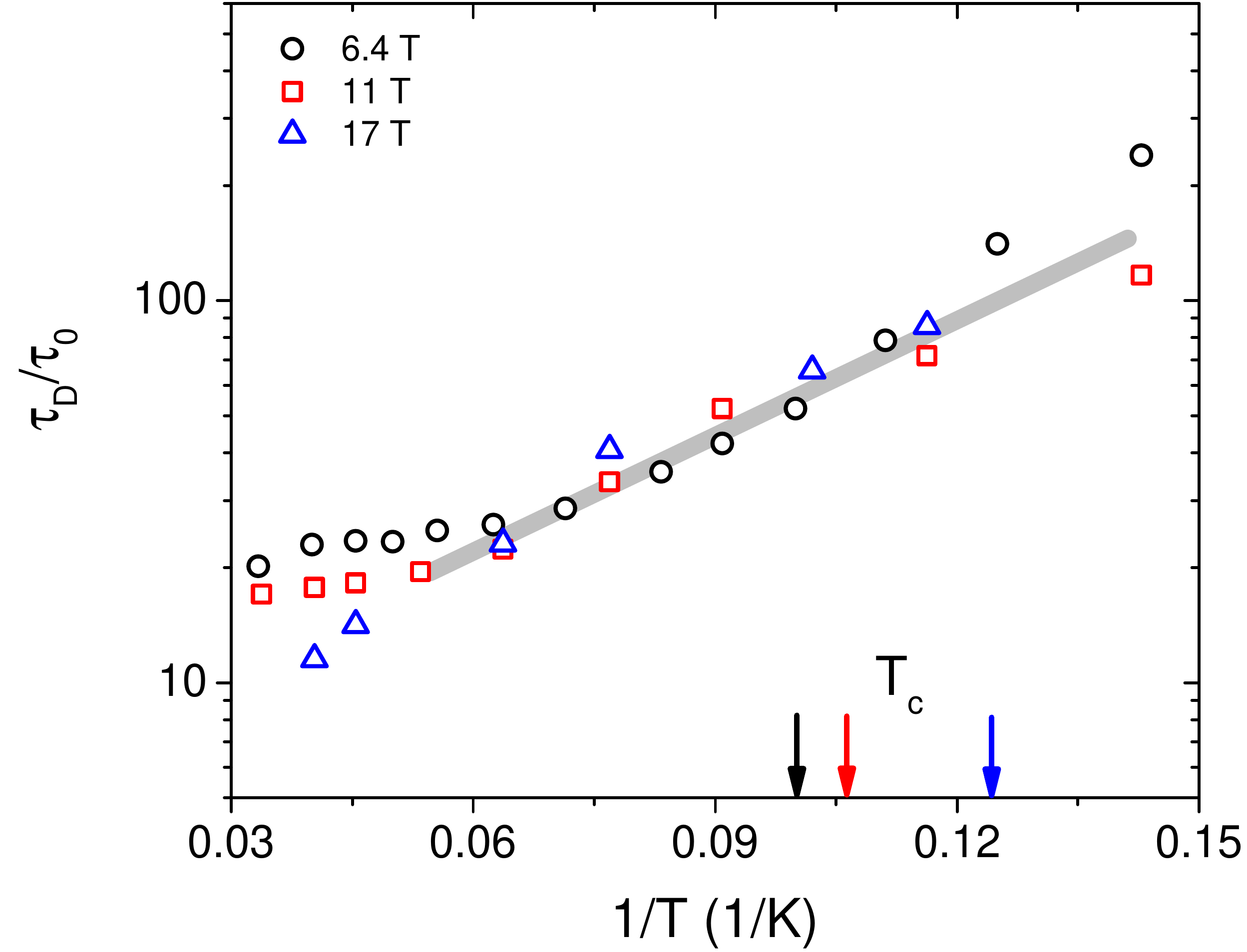}
    \caption{
Diffusion time divided by $\tau_0=1-10$ ns, for
the x=10.7\% sample, at different magnetic field strengths. The shadowed line is a linear fit to the Arrhenius law,
as described in the text. The arrows indicate $T_c$. }
    \label{nFig5}
    \end{figure}

Assuming that
the periodicity of the field gradient is equal to the diffusion length, we can write for the Hahn echo
decay rate \cite{Robertson1966}:
\begin{equation}
\frac{1}{T_{2}} (T) \simeq\frac{(\pi a\Delta\nu (H,T) )^2}{ 120
D(T)}+\frac{1}{T_{2i}}
\end{equation}
where $T_{2i}$ is the intrinsic relaxation time in absence of
dynamics. $D$ is the spin-diffusion coefficient directly related
to the characteristic fluctuation time $\tau_D=a^2/D(T)$. From the raw $\Delta \nu$
 \citep{Bossoni2013}, it is then straightforward to derive $\tau_D$. The spin-diffusion time can be fitted to an Arrhenius
law $\tau_D(T)=\tau_0 e^{U/T}$ (Fig. \ref{nFig5}), with
$\tau_0=$1-100 ns. 

\begin{figure}[htbp]
    \centering
    \includegraphics[width=7.5cm,keepaspectratio]{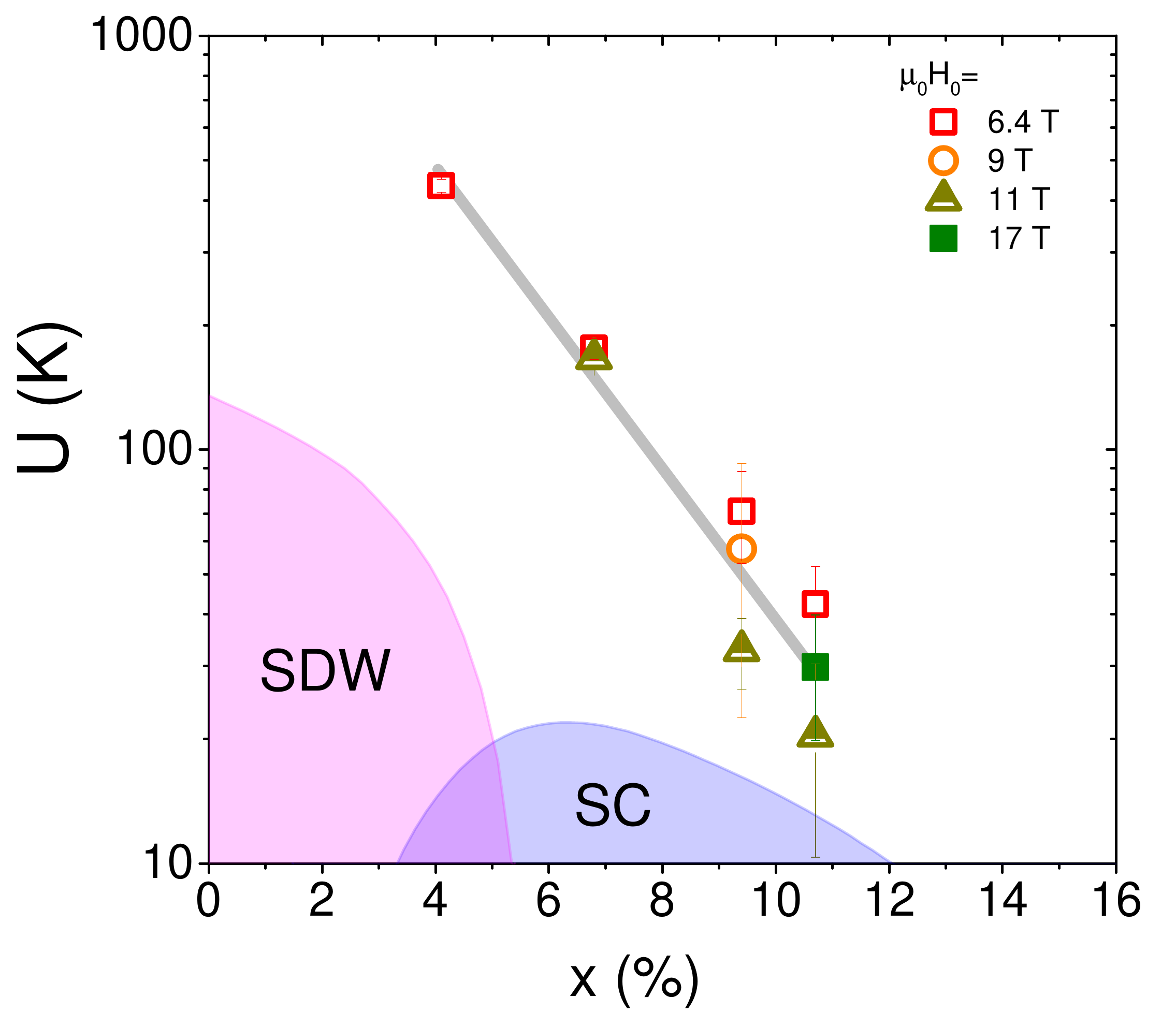}
    \caption{Activation barrier $U$ as a function of Rh content (\textit{x})
    for different magnetic fields. The energy barrier displays an exponential decrease with doping (grey line).
    The effect of the magnetic field is weak in almost all samples. The shadowed domes represent the superconducting and Spin Density Wave regions of the phase diagram.
    }
    \label{nFig6}
    \end{figure}

We found that the energy barrier $U$ decreases
nearly exponentially with Rh doping (Fig. \ref{nFig6}), and, as shown in Fig.\ref{nFig5}, it is weakly
affected by the magnetic field.   Moreover we notice that, being
$\tau_D=0.1-1$ $\mu$s, the condition of applicability of Eq. (4),
namely $\tau>>a^2/D$, is satisfied \cite{Robertson1966}. We also notice that the field dependence of $1/T_2$ observed in our previous work \cite{Bossoni2013} is here justified by Eq. 4, where the linewidth explicitly enters into the Hahn echo decay time.

\begin{figure}[htbp]
    \centering
    \includegraphics[width=8cm,keepaspectratio]{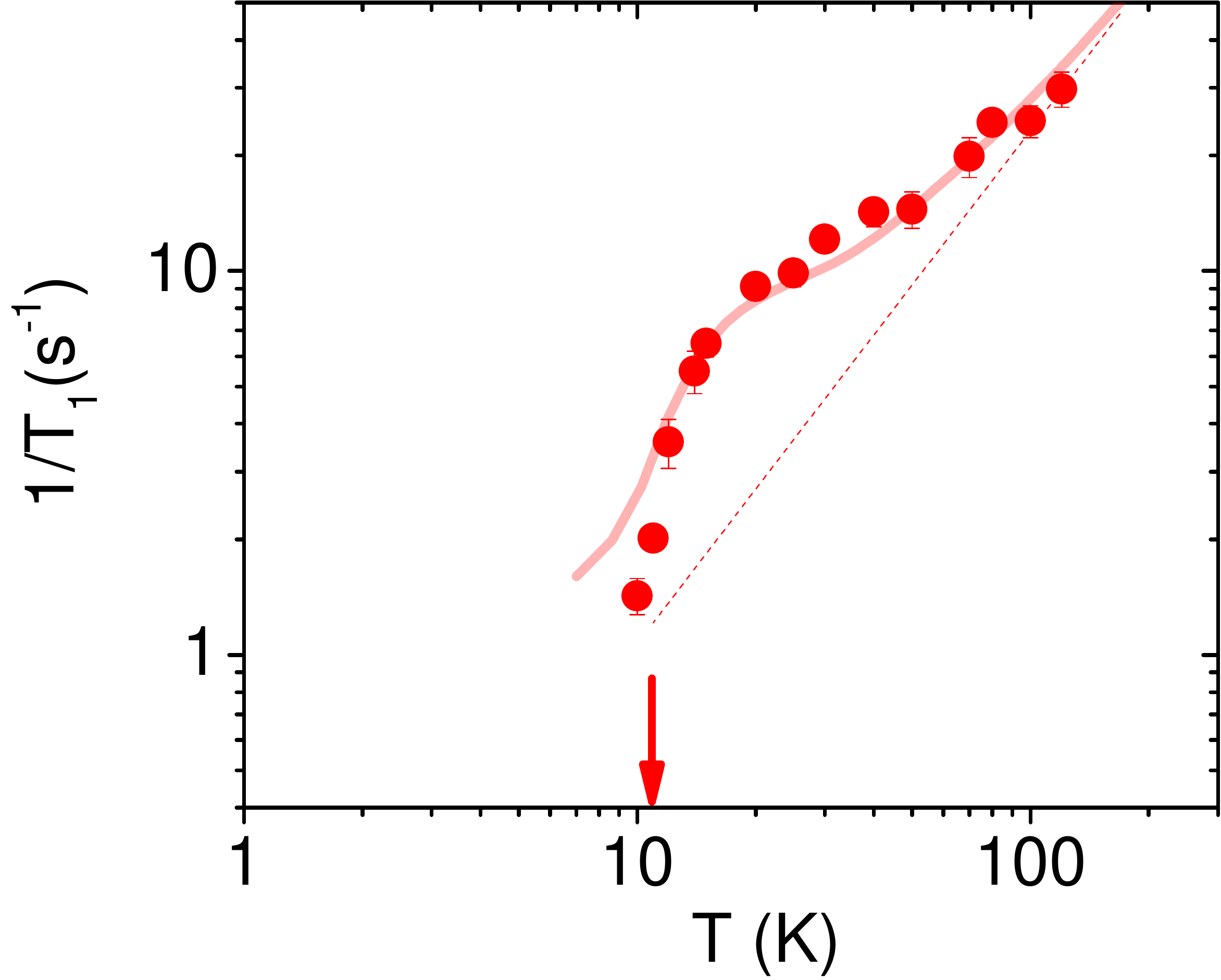}
    \caption{
Spin-lattice relaxation rate as a function of the temperature for
the $x= 9.4$ \% sample, measured with the in-plane field 
(\textbf{H} $\perp c$).    The solid line is the fit according to
Eq. 5 in the text.     The dashed line represents the
spin-lattice relaxation rate behavior for \textbf{H} $\parallel c$.}
    \label{nFig7}
    \end{figure}

The model of restricted spin diffusion can shed light also on the temperature dependence of the spin-lattice relaxation rate.
The latter slightly deviates from
linearity above $T_c$, evidencing the
presence of weak magnetic correlations
\citep{Takigawa1991,Korringa}. Additionally, the optimally
doped and weakly overdoped compounds display a hump in the
spin-lattice relaxation rate, above $T_c$ \cite{Hammerath2013,Bossoni2013},
when the magnetic field is oriented in-plane ($\mathbf{H} \perp
c$). If we assume that the spin diffusion is associated with
 random fluctuating local fields, which can be
described by a correlation function $g(t)=h_0 e^{-t/\tau_D}$, the spectral density of spin fluctuations at the Larmor frequency
$\omega_L$ \cite{Slichter} leads to:
\begin{equation}
\frac{1}{T_1}=A\frac{(2\pi \Delta\nu )^2}{ 1+(\omega_L
\tau_D)^2}\tau_D+BT^b \label{T1}
\end{equation}
The first term corresponds to the so-called
Bloembergen-Purcell-Pound (BPP) model \cite{BPP1,BPP2}, with
root mean-squared value of the transverse field equal to $ 2\pi \Delta \nu/ \gamma$
and correlation time for the field fluctuation equal to the
diffusion time $\tau_D$. The second term in Eq. 5 accounts for the
weakly correlated electron spin dynamics and for deviations from
the Korringa law. The fit in Fig. \ref{nFig7} is obtained from
three parameters, $\tau_D$, \textit{B} and \textit{b}, where the
latter two can be fixed from the high temperature regime. Despite
its simplicity, this model captures the essential features of the
experimental results, except around $T_c$, owing to the opening of
the superconducting gap. Furthermore, the
fit in Fig. \ref{nFig7} returns an energy barrier of $U\simeq 50$
K which agrees with that derived from the analysis of $1/T_2$ (see
Fig. \ref{nFig6} for comparison). Therefore, the hump in $1/T_1$
can be attributed to the very same diffusive-like dynamics that
give rise to the enhancement in $1/T_2$.

Analogous activated behavior with similar $U$ values have been
reported in LaFeAsO$_{1-x}$F$_x$
\cite{Hammerath2013,Hammerath2015}. However, in contrast to
LaFeAsO$_{1-x}$F$_x$, here the energy barrier is significantly
doping dependent, with a marked decrease upon increasing the Rh
content (Fig. \ref{nFig6}). A natural question that arises is whether this
energy scale goes to zero at a finite doping level and whether
this doping defines a quantum critical point (QCP)
\cite{Zhou2013}. We cannot fully address this question here, but
the close values of $U$ for the samples with 9.4\% and 10.7\%
samples do not point towards a QCP associated with these dynamics,
which seem rather to persist in the overdoped
regime and slowly fade away with superconductivity.

We found that the low-temperature increase of $^{75}$As $1/T_2$ in
Ba(Fe$_{1-x}$Rh$_x$)$_2$As$_2$ is not associated with an increase
in the amplitude of the spin fluctuations, as it is the case for
$^{63}$Cu $1/T_2$ in the normal phase of superconducting cuprates 
\cite{Takigawa1994}. Here, $1/T_2$ increases mostly
due to a slowing down of the dynamics  to the MHz range. These
slow dynamics are also evidenced by a field dependent hump in
$1/T_1$ (Fig. \ref{nFig7}) \cite{Hammerath2013}, while
in the cuprates $1/T_1$ is dominated by high frequency correlated
spin dynamics yielding only a weak, if any,  magnetic field
dependence of $1/T_1$ at $T> T_c$ \cite{Gorny2001,Vesna2002}.
The only case in cuprates that bears some resemblance to our data is the increase of $1/T_2$ triggered below the onset of charge order \cite{Wu2011,Hunt1999,Zaanen}. However, no evidence of charge order has been found in pnictides. Moreover, the experimental evidence that in
Ba(Fe$_{1-x}$Rh$_x$)$_2$As$_2$ these low frequency fluctuations
extend from the underdoped to the overdoped regime rather suggests
that the normal phase of iron-based superconductors is
characterized by unconventional excitations which are absent in
the cuprates.

Even if electron-doped iron based superconductors are generally
considered as itinerant systems with moderate electron
correlations, the $J_1-J_2$ model has been shown to effectively
provide an insightful approach to describe some of their magnetic
properties. In particular, in the prototypes of $J_1-J_2$ model,
it has been observed that, similarly to what we found here, slow
dynamics develop for $T< J_1+J_2$, at frequencies several orders
of magnitude below $k_B (J_1+J_2)/\hbar$ \cite{Melzi}. This has
been ascribed to activated fluctuations of domain walls separating
regions with $(\pi/a,0)$ and $(0,\pi/a)$ correlations. The
correspondent energy barrier agrees with the theoretical
prediction by Chandra, Coleman and Larkin \cite{Chandra}. More
recently, Mazin and Johannes \cite{Mazin2009} have suggested that
such low frequency domain wall excitations should be present also
in the iron-based superconductors. Therefore, it is likely that the
very slow fluctuations seen here are related to the dynamics of
domain walls separating nematic domains with perpendicular
magnetic wave-vectors. Within that framework, the energy barrier
should scale with the square of the in-plane electron spin
correlation length \cite{Chandra} and the decrease of $U$ would
indicate a decrease of electron correlations with electron doping
\cite{Medici}.

While evidence for spin nematic and orbital nematic fluctuations,
even well above the ordering temperature, have been reported in
the underdoped regime of iron-based superconductors \cite{Millis2014,Fu2012,Iye,Z2016}, no
clear evidence for the persistence of slow fluctuations driven by
nematicty has been presented for the overdoped iron-based
superconductors. It is interesting to notice that the vanishing of
the spin fluctuations probed by $1/T_2$ is accompanied by a
decrease in the amplitude of charge fluctuations of nematic
character probed by inelastic Raman scattering \cite{Gallais2013},
as well as by a decrease of the orbital anisotropy \cite{Sonobe}.
The persistence of nematic fluctuations in the overdoped regime
\cite{Gallais2013} appears consistent with our finding of slow
fluctuations remaining well above optimal doping.

\section{Conclusions}

By measuring the spin echo decay rate with different
pulse sequences, we have evidenced the presence of low-frequency
fluctuations developing in the normal phase of
Ba(Fe$_{1-x}$Rh$_x$)$_2$As$_2$ iron-based superconductors. The
comparison between $1/T_{2cpmg}$ and $1/T_2$ has suggested the presence of restricted
spin diffusive dynamics. Within this framework, the behavior of
$1/T_2$ and of $1/T_1$ can be analyzed consistently and the
fluctuations can be described by an activated correlation time with an
energy barrier exponentially decreasing with Rh-doping. Our
results point out that very slow spin dynamics persist into the
overdoped regime and could be tentatively associated with domain
walls fluctuations. These dynamics, which are an indirect
consequence of the presence of nematic correlations, are likely to
be observed in all electron doped iron-based superconductors.

\section*{acknowledgments}
R. Fernandes, Y. Gallais and R. Zhou are thanked for useful
discussions. We acknowledge Mladen Horvati\'{c} for technical
assistance and useful discussion. Jelmer Wagenaar, Martin de Wit
and Tjerk Oosterkamp are thanked for critical revision of the
manuscript. Work done at Ames Laboratory (P.C.C.) and Northwestern
University (W.P.H.) was supported by the U.S. Department of
Energy, Office of Basic Energy Science, Division of Materials
Sciences and Engineering (at Northwestern University, award No.
DE-FG02-05ER46248). The research was performed at the Ames
Laboratory. Ames Laboratory is operated for the U.S. Department of
Energy by Iowa State University under Contract No. DE-
AC02-07CH11358. This work was supported by MIUR- PRIN2012 Project
No. 2012X3YFZ2

\bibliography{ref}

\end{document}